\documentclass[prb,twocolumn,amsmath,amssymb,superscriptaddress,showpacs]{revtex4}
\usepackage{graphicx}% Include figure files
\usepackage{dcolumn}% Align table columns on decimal point
\usepackage{bm}% bold math
\def\*#1{\underline{\strut#1}}

\begin{document}

%\preprint{APS/123-QED}

\title{Excitons and High-Order Optical Transitions in Individual Carbon Nanotubes}

\author{St\'ephane Berciaud}
\affiliation{Departments of Physics and Electrical Engineering, Columbia University, New York, NY 10027, USA}
\affiliation{Department of Chemistry, Columbia University, New York, NY 10027, USA }

\author{Christophe Voisin}
\email{christophe.voisin@lpa.ens.fr}
\affiliation{Departments of Physics and Electrical Engineering, Columbia University, New York, NY 10027, USA}
\affiliation{Laboratoire Pierre Aigrain, \'Ecole Normale Sup\'erieure, 24 rue Lhomond, 75005 Paris, France}

\author{Hugen Yan}
\affiliation{Departments of Physics and Electrical Engineering, Columbia University, New York, NY 10027, USA}
\author{Bhupesh Chandra}
\affiliation{Department of Mechanical Engineering, Columbia University, New York, NY 10027, USA }
\author{Robert Caldwell}
\affiliation{Department of Mechanical Engineering, Columbia University, New York, NY 10027, USA }
\author{Yuyao Shan}
\affiliation{Department of Mechanical Engineering, Columbia University, New York, NY 10027, USA }
\author{Louis E. Brus}
\affiliation{Department of Chemistry, Columbia University, New York, NY 10027, USA }
\author{James Hone}
\affiliation{Department of Mechanical Engineering, Columbia University, New York, NY 10027, USA }
\author{Tony F. Heinz}
\email{tony.heinz@columbia.edu}
\affiliation{Departments of Physics and Electrical Engineering, Columbia University, New York, NY 10027, USA}

\begin{abstract}
We examine the excitonic nature of high-lying optical transitions in single-walled carbon nanotubes by means of Rayleigh scattering spectroscopy. A careful analysis of the principal transitions of individual semiconducting and metallic nanotubes reveals that in both cases the lineshape is consistent with an excitonic model, but not one of free-carriers. For semiconducting species, side-bands are observed at $\sim$200~meV above the third and fourth optical transitions. These features are ascribed to exciton-phonon bound states. Such side-bands are not apparent for metallic nanotubes, as expected from the reduced strength of excitonic interactions in these systems.
\end{abstract}

\pacs{78.67.Ch, 71.35.-y, 73.22.-f, 78.35.+c, 78.30.-j}
\maketitle

Excitonic effects play a central role in defining the optical properties of quasi-one dimensional (1D) materials, such as single-walled carbon nanotubes (SWNTs) \cite{book}. In the case of semiconducting species ($S$-SWNTs), confinement leads to many-body and excitonic corrections that are a significant fraction of the single particle bandgap. These effects have been predicted theoretically \cite{capaz, perebeinos-PRL, kane-mele} and validated experimentally by means of various photoluminescence excitation (PLE) schemes \cite{heinz-2photons, maultzsch-exciton, lefebvre-PRL}. Remarkably, as a consequence of the reduced screening of Coulomb interactions in 1D systems, excitonic signatures were also observed in the absorption spectra of individual metallic species ($M$-SWNTs) \cite{wang-abs}. 

Although the aforementioned experiments have unambiguously established the excitonic character of the two lowest optical transitions in $S$-SWNTs, recent spectroscopic analysis \cite{araujo, michel} reported that the energies of the third ($S_{33}$) and fourth ($S_{44}$) transitions follow a different scaling law as a function of the diameter than that of the first ($S_{11}$) and second ($S_{22}$) ones \cite{kane-mele}. It was therefore suggested that the $S_{33}$ and $S_{44}$ transitions may not arise from excitons, but rather from free-carrier contributions \cite{araujo, michel}.
In order to investigate the role of excitonic effects on high-order optical transitions, we have studied the optical response of isolated suspended SWNTs. This geometry allowed us to minimize environmental perturbations and circumvent the extrinsic effects inherent to ensemble measurements. 

In this letter, we present a detailed lineshape analysis of the Rayleigh scattering spectra of high-order transitions in structure-assigned SWNTs. For both $S$-SWNTs and $M$-SWNTs, we find that the main Rayleigh lines fit very well to an excitonic model, while a free-carrier model fails to reproduce the spectroscopic features. In addition, we report on the systematic observation of well-defined phonon side-bands (PSBs) in the Rayleigh spectra of $S_{33}$ and $S_{44}$ transitions in semiconducting species. These features have been predicted to arise from exciton-phonon bound states \cite{perebeinos-PRL}. They are recognized as additional evidence for the excitonic character of the lower optical transitions ($S_{11}$) in $S$-SWNTs \cite{plentz, berciaud, murakami, torrens}.
Such side-bands are not apparent near the $M_{22}$ Rayleigh feature in metallic species. This is consistent with the reduced strength of excitonic interactions in metallic nanotubes \cite{wang-abs}.
Our findings provide direct evidence for an excitonic description of high-order optical transitions in SWNTs and represent an important step towards resolving uncertainties about the nature of these transitions and the strength of exciton-phonon coupling in $M$-SWNTs \cite{wang-abs,berciaud,zeng}.

Rayleigh scattering spectroscopy is a powerful technique for studying the optical transitions in SWNTs \cite{book, sfeir-TEM}. For SWNTs with diameters of ~ 2~nm, the $S_{33}$, $S_{44}$ and $M_{22}$ transitions lie in the visible spectral range and are thus conveniently observed. Recent advances in supercontinuum generation have led to improved signal-to-noise ratios. Thus, whereas previous Rayleigh studies relied mainly on the spectral position of the resonances, we are now able to conduct reliable lineshape analyses and gain valuable information about underlying microscopic mechanisms. Rayleigh spectra of individual suspended nanotubes were obtained using a commercial supercontinuum laser system. The raw spectra were corrected for the supercontinuum spectrum and for the $\omega^{3}$ systematic dependence of the scattering cross-section expected for an infinite cylinder (for details, see \cite{book,sfeir-TEM}).

A hybrid experimental/theoretical scheme was developed for the assignment of nanotube structure. First, for some key nanotubes both the Rayleigh spectra and electron diffraction patterns were measured, with the latter providing direct determination of the $(n, m)$ chiral indices \cite{sfeir-TEM}. Then, for other nanotube structures, the assignments were made not based on absolute values of predicted transition energies \cite{popov}, but rather on the family patterns \cite{sfeir-TEM, araujo, michel, popov}, with the energies adjusted to match the established experimental reference points. The assignments were further confirmed by Raman scattering spectra.  These measurements provided important information on the nanotube diameter (from the radial breathing mode (RBM) frequency) \cite{sauvajol-TEM, dresselhaus-review}  and its semiconducting or metallic nature (from the G mode profile) \cite{sauvajol-TEM, dresselhaus-review, wu}.

\begin{figure}[!tb]
\includegraphics[scale=1.3]{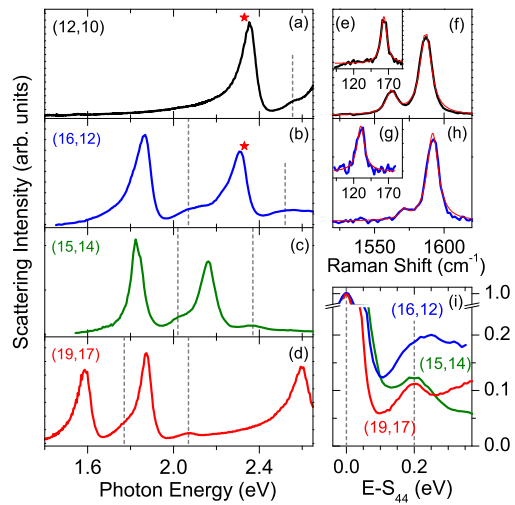}
\caption{(color online). (a-d) Rayleigh scattering spectra of individual suspended $S$-SWNTs ((mod$(n-m),3)=1,2$) with increasing diameters. The scattering intensities have been normalized by the cube of the photon energy. %The $(n,m)$ indices assigned to each nanotube are indicated. 
The vertical dotted lines highlight the $\sim$200~meV blue-shifted phonon side-bands. The stars in the Rayleigh spectra (a-b) indicate the laser energy for the Raman measurements. (e-h) Raman data for a (12,10) and a (16,12) SWNT. The RBM frequencies of (e) 163~cm$^{-1}$ and (h) 130~cm$^{-1}$ yield diameters of 1.50~nm and 1.98~nm, respectively \cite{sauvajol-TEM}. The semiconducting nature of the nanotube is further confirmed by the narrow and bimodal Raman G feature \cite{dresselhaus-review}. The Raman spectra have been fit to Lorentzian profiles. (i) Close-up of the PSB associated with the $S_{44}$ line for 3 different SWNTs. The spectra have been shifted horizontally and normalized for clarity.}
\label{fig:SB-SC} 
\end{figure}

We performed such all-optical structure assignments for several tens of individual nanotubes. A set of typical Rayleigh spectra for $S$-SWNTs with diameters ranging from 1.50~nm to 2.45~nm is displayed in Fig.~\ref{fig:SB-SC}(a-d), together with further Raman data (Fig.~\ref{fig:SB-SC}(e-h)). The main Rayleigh peaks correspond to the $S_{33}$ and $S_{44}$ transitions. For the largest diameter tubes (Fig.~\ref{fig:SB-SC}(d)), the $S_{55}$ line can also be observed. Interestingly, clear secondary features are visible at about 200~meV above each main Rayleigh line. In some cases, these features overlap with the low-energy tail of the adjacent transition; still a sizeable shoulder is observed.
Because of their location at 200$\pm$20~meV above the main line for all chiral indices, we assign them to PSBs.

This $\sim 200$~meV blue shift is close to the energy of the zone center LO ($\Gamma$-point, 197~meV) and near zone edge TO ($K$-point, 165~meV) optical phonon modes, which are well known to dominate the Raman G- and D- lines, respectively \cite{dresselhaus-review}. From the asymmetric location of the absorption- and emission side-bands relative to the main excitonic line, Torrens \textit{et al.} have established that PSBs arise chiefly from the mixing of $K$-momentum dark excitonic states with $K_{TO}$ phonons \cite{torrens}. Following this analysis, we show that Rayleigh spectroscopy of PSBs permits the investigation of $K$-momentum excitons for any transition (within our spectral detection range) and any kind of $S$-SWNT. 
Considering the energy of the $K_{TO}$ phonons, we estimate a splitting of roughly 35~meV between the bright and the $K$-momentum dark excitonic levels of both $S_{33}$ and $S_{44}$ transitions. In the diameter range investigated here, we conclude that the dependence of the excitonic manifolds on the diameter and chiral angle is too weak to be resolved.
Surprisingly, the
splittings we deduce from the PSBs of $S_{33}$ and $S_{44}$ closely match the values reported thus far for the $S_{11}$ and $S_{22}$ exciton manifolds in narrower SWNTs (0.75-1.05~nm) wrapped with surfactants or polymers \cite{plentz, torrens, berciaud, murakami}. We believe that this agreement is fortuitous and results from the $\kappa^{-1.4}d^{-2}$ scaling of $\Delta$ with the diameter and the dielectric constant $\kappa$ of the surrounding medium \cite{capaz}.

%----------------------

The full-width at half maximum (FWHM) of the side-bands is usually on the order of 60-100~meV, comparable with the widths of the main Rayleigh lines. 
In rare instances, we observed significantly broader PSBs (see Fig.~\ref{fig:SB-SC}(i)), including higher-energy features located 250-300~meV from the main excitonic lines (not shown). This energy range is similar to the experimental and theoretical estimates of the exciton binding energy for free-standing nanotubes in our diameter range \cite{lefebvre-PRL,capaz}. This suggests that these complex side-bands may result from a combination of PSBs and excited excitonic or continuum states. We note that excited states of the $S_{11}$ manifold have directly been observed in PLE experiments on individual SWNTs \cite{lefebvre-PRL}, but remain rather elusive in individual nanotube Rayleigh and absorption spectra \cite{wang-abs, berciaud} and ensemble samples \cite {plentz,torrens}. This presumably reflects the relative sensitivities of measurements.

The integrated intensities of the PSBs relative to that of the main lines range from 2\%-10\% (Fig.~\ref{fig:SB-SC}(i)), with no obvious dependence on the diameter or chiral angle. Theoretical studies have shown that the strengths of the PSBs are related to the joint density of states of the exciton-phonon quasi-particle. Since the optical phonon branches are weakly dispersive, this density of states is driven by the dispersion the $K$-momentum exciton, \textit{i.e.} by its effective mass \cite{torrens}.
However, no clear correlation was found between the intensities of the PSBs in our Rayleigh spectra and computed values of the effective mass of excitons \cite{popov}.

In contrast to absorption, which only probes the imaginary part of the SWNT dielectric susceptibility $\chi$, the Rayleigh scattering cross section is proportional to $\omega^3 |\chi(\omega)|^2$, \textit{i.e.} it includes response of both parts of $\chi$. Therefore the intensity of a PSB can be enhanced if it overlaps spectrally with non-resonant contributions to $\chi$, \textit{e.g.} from the tails of neighboring resonances, or the onset of free-carrier transitions. A simple calculation using Lorentzian profiles for both the excitonic and PSB lines shows that for an intensity ratio of 10\% in absorption \cite{torrens, berciaud}, the spectral weight of the PSB can reach 7\% in the Rayleigh spectrum, in agreement with experimental observations. Since the non-resonant contributions to the dielectric constant are known to depend on $(n,m)$, this may lead to enhanced PSBs for some specific chiral families.

Additional insight into the nature of high-energy transitions in SWNTs can be gained by carefully inspecting lineshapes in the Rayleigh spectra. In absorption (or PLE) spectroscopy, excitonic transitions are expected to give rise to Lorentzian line shapes, whereas free-carrier absorption leads to van Hove profiles in the case of quasi-1D materials. We have attempted to fit the experimental Rayleigh data to excitonic and a free-carrier models. The complex susceptibility of the exciton is described by a Lorentzian profile $\chi(\omega)\varpropto\left[(\omega_0-\omega)-i\gamma/2\right]^{-1}$ where $\omega_0$ is the resonance frequency and $\gamma$ a phenomenological width. For the free-carrier model, we compute the joint density of states of two symmetric one-dimensional parabolic bands (of the same effective mass) where each state has a natural width $\gamma$ and the matrix element is taken as constant near  the bottom of the band \cite{popov}. We obtain $\chi_{2}(\omega) \varpropto \frac{\omega_p^2}{\omega^2} \frac{\sqrt{\eta+\sqrt{1+\eta^2}}}{\sqrt{1+\eta^2}}$, where $\eta= \frac{\omega-\omega_0}{\gamma/2}$ and $\omega_p$ is the plasma frequency. The real part $\chi_{1}(\omega)$ is obtained from a Kramers-Kronig transform.

\begin{figure} [!tb]
\includegraphics[scale=1.25]{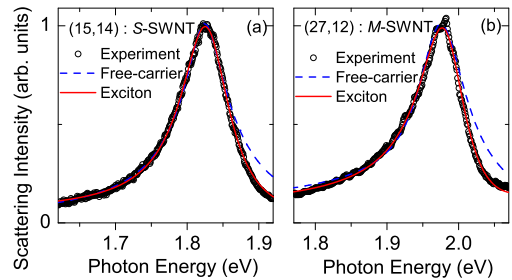}
\caption{(color online). Rayleigh scattering spectra for the $S_{33}$ transition of a (15,14) semiconducting nanotube (open circles) (a) and for the $M_{22}^-$ transition of a (27,12) metallic nanotube (b). Best fit of the profile to the free-carrier (blue dashed line) and excitonic (red solid line) models. In both cases the excitonic model is more appropriate.}
\label{fig:lineshape} 
\end{figure}

Fig.~\ref{fig:lineshape}(a) displays the Rayleigh profile of the $S_{33}$ transition in a (15,14) $S$-SWNT. Also shown are best fits to the exciton and the free-carrier models. In both cases a spectrally flat (non-resonant) component of the order of 0.2 was added to the real part of the dielectric constant. This non-resonant contribution is essential to account for the observed asymmetry of the feature. The 80~meV experimental FWHM was accounted for by using a phenomenological broadening of $\gamma$=56~meV in the free carrier model and 84~meV in the exciton model.

The excitonic model is found to provide excellent agreement with our data.
In contrast, the free-carrier model leads to poorer fits on the blue side of the line, where a systematic overestimate of the signal is observed. This reflects the asymmetric lineshape of van Hove singularities, which leads to an enhanced absorption (and modulus of the dielectric constant) on the blue side of the line compared to a (symmetric) Lorentzian. We tried to improve the model by incorporating neighboring lines at both higher and lower energies to account more quantitatively for the non-resonant contributions to the dielectric constant. The main result is unchanged: the free-carrier model always overestimates the signal on the blue side of the line and does not reproduce our data. 
This result and the presence of PSBs in the Rayleigh spectra support an excitonic description of third- and fourth-order excitations in $S$-SWNTs. 

A further interesting point concerns the linewidth of the $S_{33}$ and $S_{44}$ transitions compared with those of the $S_{11}$ and $S_{22}$ transitions. In the case of suspended SWNTs, Lorentzian fits lead to FWHMs of $\sim$10~meV and $\sim$30-50~meV for $S_{11}$ and $S_{22}$ transitions respectively \cite{lefebvre-PRL}, whereas our excitonic fits of $S_{33}$ and $S_{44}$ transitions consistently give a broadening of 80 to 90~meV.
We attribute the increased width of the $S_{33}$ and $S_{44}$ transitions to the same effect as that leading to broadening of the $S_{22}$ transition, namely, rapid relaxation into lower-lying excited electronic states.
Still, the broadening of high-energy transitions remains nearly one order of magnitude smaller than the exciton binding energy \cite{lefebvre-PRL,capaz}, consistent with the observed stability of $S_{33}$ and $S_{44}$ excitons. This broadening is compatible with a lower bound given by the lifetime of the high-energy excitons. This latter quantity can be examined by time-resolved spectroscopy and is found to be of a few tens of fs (\textit{i.e.} about 50~meV) \cite{lauret-PRL}.

\begin{figure}[!tb]
\includegraphics[scale=1.3]{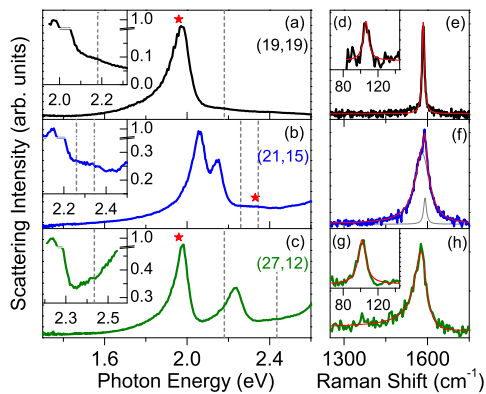}
\caption{(color online). (a-c) Rayleigh scattering spectra of individual suspended $M$-SWNTs ((mod$(n-m),3)=0$). The insets show details of the spectra near the expected positions of phonon side-bands (dotted lines). The stars in the Rayleigh spectra (a-c) indicate the laser energy for the Raman measurements. (d-h) Raman data. The RBM frequencies of (d) 106~cm$^{-1}$ and (g) 101~cm$^{-1}$ yield diameters of 2.60~nm and 2.75~nm, respectively \cite{sauvajol-TEM}. (e, f, h) G mode spectra showing chirality dependent electron-phonon coupling, as expected for chiral metallic nanotubes \cite{wu}. Spectra in (d,e,g) are fit to Lorentzian profiles, while the fits in (f,h) are based on Fano profiles.}
\label{fig:metal} 
\end{figure}

Fig.~\ref{fig:metal} displays spectra of metallic nanotubes. Chiral $M$-SWNTs are unambiguously identified through the existence of a broad Fano profile in their Raman G-mode spectra (Fig.~\ref{fig:metal}(f,h)) \cite{wu}. The high-symmetry armchair SWNT (Fig.~\ref{fig:metal}(a,d,e)) exhibits a single $M_{22}$ peak around 2~eV, together with a low RBM frequency (106~cm$^{-1}$) and a single narrow G-mode feature \cite{wu}. Phonon side-bands are not apparent in the Rayleigh spectra of the $M$-SWNTs (Fig.~\ref{fig:metal}(a,b)). Given our sensitivity, we estimate that their intensity is $<0.5\%$ of that of the $M_{22}$ lines, \textit{i.e.} at least one order of magnitude weaker than in semiconducting species. Instead of well-defined PSBs, we observe a rather strong high-energy tail\cite{comment}. This feature, particularly visible for armchair SWNTs (Fig.~\ref{fig:metal}(a)), is attributed to the existence of continuous band-to-band transitions. 
Nevertheless, the lineshapes of the main features in the Rayleigh spectra are similar to the ones of $S$-SWNTs and thus are best fit to excitonic models (Fig.~\ref{fig:lineshape}(b)).

These observations support the existence of weakly bound excitons in metallic nanotubes, in agreement with recent observations in absorption spectroscopy \cite{wang-abs}. 
The existence of excitons in metallic nanotubes is in strong contrast with the phenomenology of bulk metals and is a typical signature of 1D-confinement and the corresponding importance of Coulomb interactions. The binding energy of such excitons was estimated, both theoretically and experimentally, to be a few tens of meV \cite{wang-abs}. This is an order of magnitude less than for semiconducting nanotubes, but still significant compared to the exciton binding energies in typical 2-dimensional or 3-dimensional semiconductors.

In summary, we have carried out a careful analysis of high-order optical transitions in semiconducting and metallic carbon nanotubes of defined chiral index. These transitions are much better described by an excitonic model than by a free-carrier one. Furthermore, in semiconducting species, the third and fourth transitions systematically exhibit strong phonon side-bands, which are understood as signatures of optically forbidden states in the excitonic manifolds. The weakness of such side-bands in metallic nanotubes reflects the stronger screening of the electron-hole interaction. Thus, phonon side-bands provide a useful optical criterion for distinguishing semiconducting nanotubes from metallic ones.
We conclude that higher-lying energy transitions in semiconducting carbon nanotubes arise from excitons. This result and the observation of a dominant excitonic component in the optical transitions in metallic nanotubes highlight the unusual properties of quasi-1D materials like nanotubes. 

SB and CV contributed equally to this work. We thank Yuhei Miyauchi for informative discussions.
We acknowledge support from the Nanoscale Science and Engineering Initiative of the NSF under grants ECS-05-07111 and CHE-0117752, the US DOE under grant DE FG02 98ER 1486, and from the Intel Corporation.


\begin{thebibliography}{} 


\bibitem{book}  Carbon Nanotubes: Advanced Topics in the Synthesis, Structure, Properties and Applications, A. Jorio, M. S. Dresselhaus, G. Dresselhaus, eds. (Springer, 2007).

\bibitem{capaz} R.B. Capaz \textit{et al.}, Phys. Rev. B. {\bf 74}, 121401(R) (2006).


\bibitem{kane-mele} C.L. Kane and E.J. Mele, Phys. Rev. Lett. {\bf 93}, 197402 (2004).

\bibitem{perebeinos-PRL} V. Perebeinos \textit{et al.}, Phys. Rev. Lett. {\bf 94} 027402 (2005).


\bibitem{heinz-2photons} F. Wang \textit{et al.}, Science {\bf 308}, 838 (2005). 

\bibitem{maultzsch-exciton} J. Maultzsch \textit{et al.}, Phys. Rev. B, {\bf 72}, 241402(R) (2005).

\bibitem{lefebvre-PRL} J. Lefebvre and P. Finnie, Phys. Rev. Lett. {\bf 98}, 167406 (2007).

\bibitem{wang-abs} F. Wang \textit{et al.}, Phys. Rev. Lett. {\bf 99}, 227401 (2007).

\bibitem{araujo} P.T. Araujo \textit{et al.}, Phys. Rev. Lett. {\bf 98}, 067401 (2007).

\bibitem{michel} T. Michel \textit{et al.}, Phys. Rev. B. {\bf 75}, 155432 (2007).

\bibitem{torrens} O. N. Torrens \textit{et al.}, Phys. Rev. Lett. {\bf 101}, 157401 (2008).

\bibitem{plentz} F. Plentz \textit{et al.}, Phys. Rev. Lett. {\bf 95}, 247401 (2005).

\bibitem{berciaud} S. Berciaud \textit{et al.}, Nano Lett. {\bf 7}, 1203 (2007).

\bibitem{murakami} Y. Murakami \textit{et al.}, Phys. Rev. B. {\bf 79}, 195407 (2009).

\bibitem{zeng} H. Zeng \textit{et al.}, Phys. Rev. Lett. {\bf 102}, 136406 (2009).

\bibitem{sfeir-TEM} M.Y. Sfeir \textit{et al.}, Science {\bf 312} 554 (2006), and references therein.

\bibitem{popov} V. N. Popov New. J. Phys. {\bf 6}, 17 (2004).

\bibitem{sauvajol-TEM} J. C. Meyer \textit{et al.}, Phys. Rev. Lett. {\bf 95}, 217401 (2005).

\bibitem{dresselhaus-review} M.S. Dresselhaus \textit{et al.}, Phys. Rep. {\bf 409}, 47 (2005).

\bibitem{wu} Y. Wu \textit{et al.}, Phys. Rev. Lett. {\bf 99}, 027402 (2007).

\bibitem{lauret-PRL} J.-S. Lauret \textit{et al.}, Phys. Rev. Lett. {\bf 90}, 057404 (2003).

\bibitem{comment} Zeng \textit{et al}. reported the observation of a strong PSB in the reflectance spectrum of an individual suspended nanotube, described as a (13,10) M-SWNT \cite{zeng}. In view of its narrow bimodal Raman G-mode feature (Fig. 1(a) in ref. \cite{zeng}), we would, however, assign it as a semiconducting nanotube. 

%---------------
%SAME references but with all authors listed
%\bibitem{book}  Carbon Nanotubes: Advanced Topics in the Synthesis, Structure, Properties and Applications, A. Jorio, M. S. Dresselhaus, G. Dresselhaus, eds. (Springer, Berlin, 2007)
%\bibitem{capaz} R.B. Capaz, C.D. Spataru, S. Ismail-Beigi and S.G. Louie, Phys. Rev. B. {\bf 74}, 121401(R) (2006).
%\bibitem{kane-mele} C.L. Kane and E.J. Mele, Phys. Rev. Lett. {\bf 93}, 197402 (2004).
%\bibitem{perebeinos-PRL} V. Perebeinos, J. Tersoff and P. Avouris, Phys. Rev. Lett. {\bf 94} 027402 (2005).
%\bibitem{heinz-2photons} F. Wang, G. Dukovic, L.E. Brus, T.F. Heinz, Science {\bf 308}, 838 (2005). 
%\bibitem{maultzsch-exciton} J. Maultzsch, R. Pomraenke, S. Reich, E. Chang, D. Prezzi, A. Ruini, E. Molinari, M.S. Strano, C. Thomsen and C. Lienau, Phys. Rev. B, {\bf 72}, 241402(R) (2005).
%\bibitem{lefebvre-PRL} J. Lefebvre and P. Finnie, Phys. Rev. Lett. {\bf 98}, 167406 (2007).
%\bibitem{wang-abs} F. Wang, D. J. Cho, B. Kessler, J. Deslippe, P. J. Schuck, Steven G. Louie, A. Zettl, T.F. Heinz and Y.R. Shen, Phys. Rev. Lett. {\bf 99}, 227401 (2007).
%\bibitem{araujo} P.T. Araujo, S. K. Doorn, S. Kilina, S. Tretiak, E. Einarsson, S. Maruyama, H. Chacham, M. A. Pimenta and A. Jorio, Phys. Rev. Lett. {\bf 98}, 067401 (2007).
%\bibitem{michel} T. Michel, M. Paillet, J.C. Meyer, V.N. Popov, L. Henrard and J-L. Sauvajol, Phys. Rev. B. {\bf 75}, 155432 (2007).
%\bibitem{torrens} O. N. Torrens, M. Zheng and J.M. Kikkawa, Phys. Rev. Lett. {\bf 101}, 157401 (2008).
%\bibitem{plentz} F. Plentz, H.B. Ribeiro, A. Jorio, M.S. Strano and M.A. Pimenta, Phys. Rev. Lett. {\bf 95}, 247401 (2005).
%\bibitem{berciaud} S. Berciaud, L. Cognet, P. Poulin, R.B. Weisman and B. Lounis, Nano Lett. {\bf 7}, 1203 (2007).
%\bibitem{murakami} Y. Murakami, B. Lu, S. Kazaoui, N. Minami, T. Okubo, and S. Maruyama, Phys. Rev. B. {\bf 79}, 195407 (2009).
%\bibitem{zeng} H. Zeng, H. Zhao, F-C Zhang and X. Cui, Phys. Rev. Lett. {\bf 102}, 136406 (2009).
%\bibitem{sfeir-TEM} M.Y. Sfeir, T. Beetz, F. Wang, L. Huang, X.M.H. Huang, M. Huang, J. Hone, S. O'Brien, J. A. Misewich, T. F. Heinz, L. Wu, Y. Zhu and L. E. Brus, Science {\bf 312} 554 (2006), and references therein.
%\bibitem{popov} V. N. Popov New. J. Phys. {\bf 6}, 17 (2004).
%\bibitem{sauvajol-TEM} J. C. Meyer, M. Paillet, T. Michel, A. Mor\'eac, A. Neumann, G. S. Duesberg, S. Roth, and J.-L. Sauvajol, Phys. Rev. Lett. {\bf 95}, 217401 (2005).
%\bibitem{wu} Y. Wu, J. Maultzsch, E. Knoesel, B. Chandra, M. Huang, M.Y. Sfeir, L.E. Brus, J. Hone, and T.F. Heinz, Phys. Rev. Lett. {\bf 99}, 027402 (2007).
%\bibitem{dresselhaus-review} M.S. Dresselhaus, G. Dresselhaus, R. Saito and A. Jorio, Phys. Rep. {\bf 409}, 47 (2005).
%\bibitem{lauret-PRL} J.-S. Lauret, C. Voisin, G. Cassabois, C. Delalande, P. Roussignol, O. Jost and L. Capes, Phys. Rev. Lett. {\bf 90}, 057404 (2003).
%------------------------




\end{thebibliography}
\end{document}